# Direct Measurement of the Elastohydrodynamic Lift Force at the Nanoscale

Zaicheng Zhang[1], Vincent Bertin[1,2], Muhammad Arshad[1], Elie Raphaël[2], Thomas Salez[1,3], and Abdelhamid Maali[1, *]

[1] Univ. Bordeaux, CNRS, LOMA, UMR 5798, F-33405, Talence, France.
[2] UMR CNRS Gulliver 7083, ESPCI Paris, PSL Research University, 75005 Paris, France.
[3] Global Station for Soft Matter, Global Institution for Collaborative Research and Education, Hokkaido University, Sapporo, Hokkaido 060-0808, Japan.

**Abstract**

We present the first direct measurement of the elastohydrodynamic lift force acting on a sphere moving within a viscous liquid, near and along a soft substrate under nanometric confinement. Using atomic force microscopy, the lift force is probed as a function of the gap size, for various driving velocities, viscosities, and stiffnesses. The force increases as the gap is reduced and shows a saturation at small gap. The results are in excellent agreement with scaling arguments and a quantitative model developed from the soft lubrication theory, in linear elasticity, and for small compliances. For larger compliances, or equivalently for smaller confinement length scales, an empirical scaling law for the observed saturation of the lift force is given and discussed.

* Corresponding author: abdelhamid.maali@u-bordeaux.fr

Lubricated contact between deformable solids is a situation widely encountered and studied in geophysical [1], as well as industrial and engineering [2] contexts. Central to such elastohydrodynamic settings is the coupling between the local hydrodynamic pressure induced by the fluid flow and the deformation of the confining solids. Recently, such a coupling was studied for much more compliant solids and smaller length scales, in the context of soft matter in confinement and at interfaces [3]. Indeed, it might play a crucial role in the motion of various physiological and biological entities[4,5]. Furthermore, through surface force apparatus [6-9], atomic force microscopy [10-14], and optical particle tracking [15], it offers an alternative strategy for micro- and nanorheology of soft materials, with the key advantage of avoiding any solid-solid adhesive contact.

In such a soft-matter context, a novel elastohydrodynamic lift force was theoretically introduced for elastic bodies moving past each other within a fluid [16], and further explored and generalized through: the motion of vesicles along a wall [17,18], different elastic media and geometries [19-21], added effects of intermolecular interactions [22], self-similar properties of the soft lubricated contact [23], the inertial-like motion of a free particle [24], viscoelastic effects [25], an equivalent emerging torque [26], and the case of membranes [27,28]. Essentially, any symmetric rigid object moving within a viscous fluid and along a nearby soft surface is repelled from the latter by a normal force. This force arises from a symmetry breaking in the contact shape and the associated low-Reynolds-number flow, due to the elastohydrodynamic coupling introduced above. Specifically, for a non-deformable surface, and an even contact shape, the lubrication pressure field (*i.e.* the dominant hydrodynamic stress) is antisymmetric, resulting in a null net normal force. In contrast, a soft surface is deformed by the pressure field which then loses its symmetry, resulting in a finite normal force. We note that the qualitative behaviour is similar for the opposed situation of a soft object moving within a viscous fluid along a rigid surface.

Theoretical calculations show that, as the gap between the object and the soft substrate reduces, the force increases. Eventually, at very small gap, the competition between symmetry breaking and decreasing pressure leads to a saturation of the lift force [19-22].

Despite the above theoretical literature, experimental evidence for such an elastohydrodynamic lift force remains recent and scarce [29,30]. Measurements of the rising speed and the distance to a vertical wall of a bubble allowed to extract an analogous normal force acting on the bubble [31]. A qualitative observation was reported in the context of smart lubricant and elastic polyelectrolytes [32]. A study, involving the sliding of an immersed macroscopic cylinder along an inclined plane, precoated with a thin layer of gel, showed an effective reduc-tion of friction induced by the lift force [33]. Then, the optical tracking of the driven motion of a microparticle in a microfluidic channel decorated with a polymer brush revealed the potential importance of this force in biological and microscopic settings [34]. From the gravitational sedimentation of a macroscopic object along a vertical membrane under tension, another study observed an important normal drift, showing the amplification of the effect for very compliant boundaries induced by slender geometries [35]. The measurement of the shape deformation of a levitating droplet over a moving wall was also used to probe the effects of the lift force [36]. Nevertheless, while this experimental literature provides confidence in the existence of the elastohydrodynamic lift force, as well as in its importance at small scales and for biology, no direct force measurement was performed to date and the saturation at the nanoscale was not yet observed.

In this Letter, we report on the first direct measurement of the elastohydrodynamic lift force acting on a sphere moving within a viscous liquid and along a soft substrate, under nanometric confinement. Using atomic force microscopy (AFM), the lift force is probed as a function of the gap size, for various driving velocities, viscosities, and stiffnesses. The results are compared to scaling arguments and a novel quantitative model developed from the soft lubrication theory, in linear elasticity, and for small compliances. For larger compliances, a saturation of the lift force is observed and its empirical scaling law is discussed.

A schematic of the experimental setup is shown in Fig. 1. The experiment is performed using an AFM (Bruker, Bioscope) equipped with a cantilever holder (DTFML-DD-HE) that allows working in a liquid environment. We use a spherical borosilicate particle (MO-Sci Corporation) with a radius $R = 60 \pm 1$ μm and a roughness of 0.9 nm measured over a 1 μm$^2$ surface area. That sphere is glued at the end of a silicon nitride triangular shaped cantilever (DNP, Brukerafmprobes) using epoxy glue (Araldite, Bostik, Coubert). The soft samples are fixed on a multi-axis piezo-system (NanoT series, Mad City Labs), which allows: i) to control and scan the gap distance $d$ between the sphere and the sample by displacing the sample vertically; and ii) to vibrate the sample transversally at a frequency $f = \omega/(2\pi) = 25$ or 50 Hz, and with an amplitude $A$ ranging form 3.6 μm to 36 μm. Note that, the normal displacement speed 20 nm/s being much smaller than the smallest transversal velocity amplitude $A\omega = 0.57$ mm/s, the former can be neglected and a quasi-static description with respect to the normal motion is valid. Using the drainage method [37], the modified stiffness $k_c = 0.21 \pm 0.02$ N/m of the cantilever when the sphere is attached to it is determined using a rigid silicon wafer as a substrate, and for large enough gap distances ($d = 200\text{-}20000$ nm). The studied polydimethylsiloxane (PDMS) substrates are prepared as follows. First, uncrosslinked PDMS (Sylgard 184, Dow Corning) and its curing agent are mixed into three different solutions, with different mixing ratios (10:1, 20:1, 30:1). Following a preliminary degassing process, a few droplets of each solution are spin-coated on a glass substrate during a minute to get a sample of thickness in the 25-30 μm range. This is followed by an annealing step, in an oven at 50 °C and during 24 hours, in order to promote an efficient cross-linking. The measured Young's moduli $E$ of the samples (10:1), (20:1), and (30:1), are respectively: $(1455 \pm 100)$ kPa, $(600 \pm 50)$ kPa and $(293 \pm 20)$ kPa, where the Poisson ratio is fixed to $\nu = 0.5$ since crosslinked PDMS is an incompressible material to a very good approximation. At the Young's moduli and low frequencies studied here, the loss modulus of PDMS is negligible [38]. The viscous liquids employed are silicone oil and 1-decanol with dynamic viscosities $\eta = 96$ mPa·s and $14.1$ mPa·s respectively.

Using scaling arguments, the lift force acting on a sphere immersed in a viscous fluid and moving at constant velocity $V$, near and parallel to a semi-infinite incompressible elastic substrate of shear modulus $G = E/[2(1 + \nu)]$, reads [20]:

$$F_{\text{lift}} \sim \frac{\eta^2 V^2}{G} \frac{R^{5/2}}{d^{5/2}}, \tag{1}$$

in the limit of small dimensionless compliance, $\kappa = \eta V R/(G d^2) \ll 1$. Note that, in this limit, $\kappa$ corresponds to the ratio between substrate's deformation and gap distance. Note also that, due to Galilean invariance, moving the substrate at constant velocity instead of the sphere leads to the same lift force. In view of the low frequencies at which the substrate is oscillating, and since inertial effects are negligible for such a confined viscous flow, this invariance and the expression of the lift force above remain excellent approximations in our case - with the substitution $V = A\omega\sin(\omega t)$ in Eq.(1). In addition, in all experiments, the hydrodynamic radius $\sqrt{2Rd}$ being much smaller than the thickness of the soft substrate, the latter can indeed be described as semi-infinite. Interestingly, with such a periodic driving, and since the lift force depends on the squared velocity, it can be expressed as two additive components: i) a time-independent one $\sim \eta^2 A^2 \omega^2 R^{\frac{5}{2}}/(2G d^{5/2})$; and ii) a component oscillating at double fre-quency $2f$. Focusing only on the former, it is measured though a temporal average $F = \langle F_N \rangle$ of the instantaneous normal force $F_N$ recorded by AFM (see Fig.1).

Figure 2 shows the force $F$ as a function of the gap distance $d$, for rigid (silicon wafer) and soft substrates (PDMS 20:1). For the rigid case, no finite force is detected above the current nN resolution, at all distances. This is expected, since for such a hard surface (Young's modulus in the 100 GPa range), the elastohydrodynamic effects occur at gap distances much smaller than the ones typically probed here [6]. As a remark, the fact that no force – even purely hydrodynamic – is measured in this case is a direct confirmation for the validity of the quasi-static description with respect to the imposed normal motion of the sphere. In sharp contrast, for the soft case, a systematic non-zero force is measured, and observed to increase as the gap distance is reduced. Furthermore, as shown in the inset, the force asymptotically scales as $F \sim d^{-5/2}$ at large gap distances, in agreement with the prediction of Eq.(1). Interestingly, at smaller gap distances, a saturation of the lift effect is observed, as reported previously [20,33].

Having tested the asymptotic dependence of the force with the main geometrical parameter, *i.e.* the gap distance, which showed a first evidence of the lift, we now turn to the other key elastohydrodynamic parameters appearing in Eq.(1): the velocity amplitude $A\omega$, the viscosity $\eta$ of the liquid, and the shear modulus $G$ of the substrate. To test the dependences of the force with those three parameters, we introduce two dimensionless variables: the dimensionless compliance $\kappa = \eta V R / (G d^2)$, and the dimensionless force $F/F^*$ with $F^* = \eta V R^{3/2} / d^{1/2}$, where $V$ is systematically replaced by its root-mean-squared value $A\omega/\sqrt{2}$ due to the temporal averaging introduced above. In such a representation, Eq.(1) becomes: $F/F^* \sim \kappa$. In Fig.3, we thus plot $F$ as a function of $d$, and in the rescaled form, $F/F^*$ as a function of $\kappa$, for various sets of parameters: two different oscillation amplitudes (Fig.3a), two different oscillation frequencies (Fig.3b), two different viscosities (Fig.3c), and three different shear moduli (Fig.3d). In the inset of each of those panels, we first observe at small $\kappa$ that $F/F^*$ is linear in $\kappa$, and that the curves for various values of the varied parameter collapse with one another, which validates further Eq.(1). Moreover, around $\kappa \sim 1$, a deviation from the previous asymptotic behaviour is observed, leading to a maximum prior to an interesting decay at large $\kappa$. In addition, the collapse for various values of the varied parameter is maintained, indicating that even at large dimensionless compliance $\kappa$, the dimensionless force $F/F^*$ remains a function of $\kappa$ only. This suggests that the same physics, coupling lubrication flow and linear elasticity, is at play at large $\kappa$.

We now rationalize the missing prefactor in Eq.(1), and discuss further the behaviour at large $\kappa$. For the first purpose, we go beyond scaling analysis [20] and develop a model based on soft lubrication theory [19-26] for a rigid sphere translating in a viscous fluid and along a semi-infinite incompressible elastic wall. Details are provided in SI, and lead to the expression of the lift force at first order in $\kappa$:

$$F_{\text{lift}} \simeq \kappa \int d^2\boldsymbol{r}\, p_1(\boldsymbol{r}) \approx 0.416 \frac{\eta^2 V^2}{G}\left(\frac{R}{d}\right)^{5/2}, \qquad (2)$$

where $\kappa\, p_1(\boldsymbol{r})$ is the first-order lubrication pressure term, and $\boldsymbol{r} = (x, y)$ is the coordinate vector in the horizontal plan (see Fig.1a).

Equation (2) thus provides the missing prefactor of Eq. (1), allowing to go beyond scaling analysis. In order to test this prediction, we plot $F/F^*$ as a function of $\kappa$ in Fig.4, for all the experiments performed in this study. First, all the experimental data collapses on a single non-monotonic master curve, confirming further the results of Fig.3. Secondly, Eq. (2) is found to be in excellent agreement with the low-$\kappa$ part of the data, with no adjustable parameter. Finally, the behaviour at large $\kappa$ reveals the possible existence of a power law: $F/F^* \sim \kappa^{-1/4}$, equivalent to $F \sim \eta^{3/4} V^{3/4} G^{1/4} R^{5/4}$. This gap-independent empirical scaling suggests that the lift force saturates at small enough distances, in agreement with the observation made in Fig.2. Such a result [25] might tentatively be attributed to a competition between the increase of the elastohydrodynamic symmetry breaking and the decrease of the pressure magnitude due to the substrate's deformation, but further work is needed to quantify this hypothetical mechanism, and to disentangle it from potential non-stationary effects [8]. Indeed, the latter are *a priori* not negligible anymore at large $\kappa$ (see SI).

In conclusion, our results robustly demonstrate the existence and the first direct measurement of the elastohydrodynamic lift force at the nanoscale, and confirm our novel quantitative asymptotic theoretical prediction. Moreover, the latter having been developed in the framework of classical soft lubrication theory, the collapse of the data with it for various amplitudes, frequencies, viscosities, and shear moduli, allows to safely exclude artefacts from viscoelasticity, poroelasticity, or non-linear elasticity. For large compliances, or equivalently at small confinement length scales, a saturation of the lift force is observed and an empirical scaling law is discussed. In future, focusing the efforts on the resolution of the nonlinear problem at any dimensionless compliance, and including non-stationary terms associated with the driving oscillation, might help to explore further the saturation regime. We anticipate important implications of the existence of the elastohydrodynamic lift force at the nanoscale for nanoscience and biology.

**Acknowledgements**

The authors thank Yacine Amarouchene and Alois Würger for fruitful discussions. Zaicheng Zhang acknowledges financial support from the Chinese Scholarship Council. Muhammad Arshad acknowledges financial support from the higher education commission of Pakistan.

**Figures**

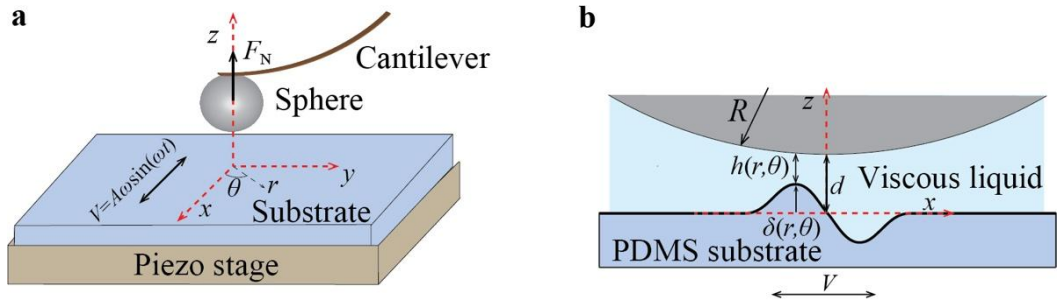

**Figure 1**. *Schematic of the experimental setup. The soft PDMS sample is fixed to a rigid piezo stage that is transversally oscillated along time t, at angular frequency ω, and with amplitude A. A rigid borosilicate sphere is glued to an AFM cantilever and placed near the substrate, with silicone oil or 1-decanol as a viscous liquid lubricant. The normal force $F_N$ exerted on the sphere, at a gap distance d from the surface, is directly measured from the deflection of the cantilever.*

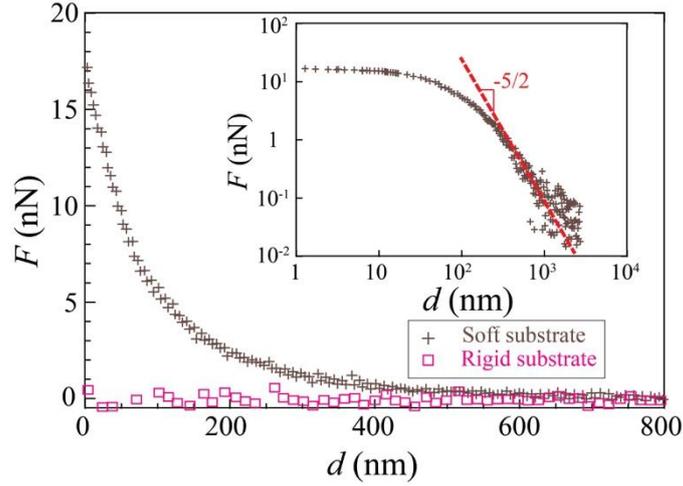

**Figure 2**. *Temporal average F of the normal force $F_N$ (see Fig.1) as a function of the gap distance d to the substrate, for both rigid (silicon wafer) and soft (PDMS 20:1) substrates. The liquid used is silicone oil with viscosity $\eta$=96 $mPa \cdot s$. The amplitude of the velocity is $A\omega = 0.57\ mm/s$. The inset shows a log-log representation of the data for the soft substrate, and the solid line therein indicates a -5/2 power law.*

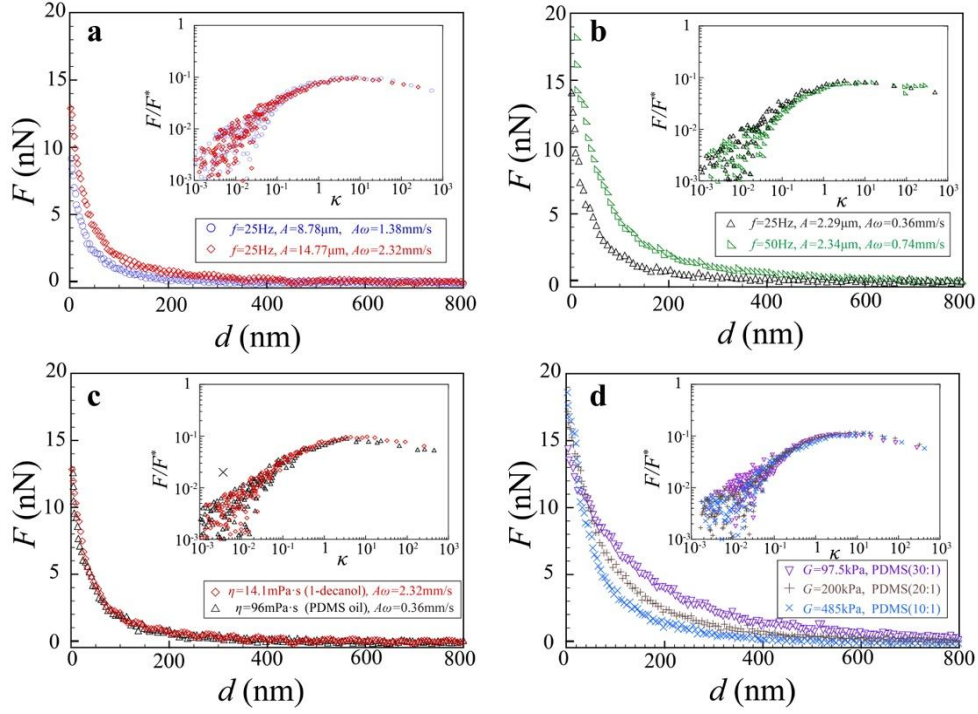

**Figure 3**. *Measured temporal-averaged force F as a function of gap distance d to the soft PDMS substrates, and (insets) dimensionless force $F/F^*$ as a function of dimensionless compliance $\kappa$ in logarithmic scales, for various sets of parameters. (a) Two different velocity amplitudes (as indicated) obtained with different oscillation amplitudes are investigated. The substrate is crosslinked PDMS (10:1), and the liquid is 1-decanol with viscosity $\eta = 14.1 \ mPa \cdot s$; (b) two different velocity amplitudes (as indicated) obtained with two different working frequencies are investigated. The substrate is crosslinked PDMS (10:1), and the liquid is silicone oil with viscosity $\eta = 96 \ mPa \cdot s$; (c) two different liquids with different associated viscosities (as indicated) are investigated. The substrate is crosslinked PDMS (10:1), and the velocity amplitudes are $A\omega = 0.36 \ mm/s$ and $A\omega = 2.32 \ mm/s$ for silicone oil ($\eta = 96 \ mPa \cdot s$) and 1-decanol ($\eta = 14.1 \ mPa \cdot s$) respectively; (d) three different shear moduli (as indicated) of the substrate are investigated. The liquid is silicone oil with viscosity $\eta = 96 \ mPa \cdot s$, and the velocity amplitude is $A\omega = 0.57 \ mm/s$.*

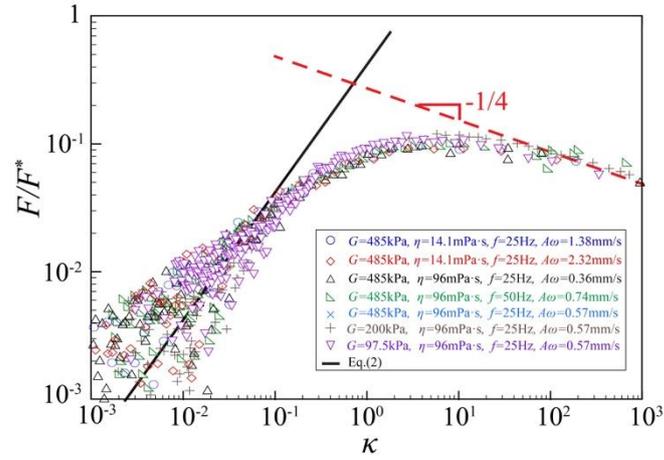

**Figure 4**. *Dimensionless force $F/F^*$ as a function of dimensionless compliance $\kappa$ (see definitions in text) in logarithmic scales, as measured from force-distance data (see Figs.2 and 3), for all the experiments performed in this study (see Fig.3). The solid line corresponds to the theoretical prediction for F at low $\kappa$, obtained from Eq.(2) where V is replaced by $A\omega/\sqrt{2}$ due to the temporal averaging step. The dashed lined indicates a -1/4 power law.*